# Extended Capability Models for Carbon Fiber Composite (CFC) Panels in the Unstructured Transmission Line Modelling (UTLM) Method

Xuesong Meng, Ana Vukovic, *Member, IEEE,* Trevor M. Benson, *Senior Member, IEEE* and Phillip Sewell, *Senior Member, IEEE*

*Abstract*— An effective model of single and multilayered thin panels, including those formed using carbon fiber composite (CFC) materials, is incorporated into the Transmission Line Modeling (TLM) method. The thin panel model is a one-dimensional (1D) one based on analytical expansions of cotangent and cosecant functions that are used to describe the admittance matrix in the frequency domain; these are then converted into the time domain by using digital filter theory and an inverse Z transform. The model, which is extended to allow for material anisotropy, is executed within 1D TLM codes. And, for the first time, the two-dimensional (2D) thin surface model is embedded in unstructured three-dimensional (3D) TLM codes. The approach is validated by using it to study some canonical structures with analytic solutions, and against results taken from the literature. It is then used to investigate shielding effectiveness of carbon fiber composite materials in a practical curved aerospace-related structure.

I. INTRODUCTION

The use of advanced materials continues to grow, especially within the aerospace sector. In particular, carbon fiber composite (CFC) materials have attracted the attention of many [1]-[6] due to their high strength-to-weight ratio and ease of fabrication. As replacements for metals, they have been used in numerous areas, such as spacecraft and aircraft structures [1], avionics systems [2], and Radio-Frequency Identification (RFID) [3]. However, due to their lower conductivity, CFC materials have a lower shielding effectiveness compared to that of metal. In order to analyze and improve the shielding effectiveness a variety of CFC materials have been developed and studied using either numerical or analytical methods.

Sarto [1] reported a FDTD model for analyzing the shielding performances of multilayered CFC panels, in which a vector fitting procedure was used to get the equations in the time domain. Holloway et al. [4] proposed three different equivalent-layer models that capture various physical aspects of this material and Rea et al. [2] investigated the shielding properties of CFC materials based on the plane wave shielding theory and effective medium theory. Mehdipour *et al.* [5] utilized an anisotropic equivalent model of each layer and combined it with the transmission matrix method (TMM) to simulate the shielding effectiveness of multilayered CFC materials.

In this paper a multi-layered thin panel model, well suited to electromagnetics-based studies of CFC structures, is implemented in The Transmission Line Modeling (TLM) method [7]. TLM is a full wave numerical method that is used for electromagnetic applications. Unlike the finite-difference time-domain (FDTD) method, it calculates the electric and magnetic fields at the same point and at the same time step ensuring the unconditional stability and appropriate for implementation of complex material properties. Materials with frequency-dependent [8], anisotropic [9] and nonlinear [10] properties have been implemented in the TLM method. The TLM method discretizes the problem in question using either structured or, more recently, unstructured [11] – [13] meshes or a combination of the two [14]. A typical requirement is that the sampling mesh size is of the order of λ/10 where λ is operating wavelength. A CFC panel is usually very thin compared to the wavelengths of interest, and so if discretized directly in the numerical method requires implementation of a very fine mesh and consequently large run times and memory storage.

In [15] we described an efficient and versatile approach for embedding a one dimensional (1D) single- or multi-layered thin sheet model in an otherwise relatively coarse grid, and applied it to the study of the reflection and transmission properties of some common photonic devices such as anti-reflection coatings and fiber Bragg gratings. In [16] the model was extended to the modeling thin single-layered curved panels within a structured two-dimensional (2D) TLM algorithm, and applied to the study of electromagnetic shielding problems involving CFC structures. It was shown that the approach was not only accurate and efficient, but also that it offered the benefit of great versatility since the panels can be composed of loss-free or lossy dielectric materials. The approach adopted in [16] was to approximate the thin panels

Manuscript received December 11, 2015. This work was supported by China Scholarship Council and Innovate UK.

The authors are with the George Green Institute for Electromagnetics Research, University of Nottingham, Nottingham, NG7 2RD, UK. (email:Xuesong.meng@nottingham.ac.uk)

using a piece-wise linearization and then to embed the linearized segments between adjacent nodes of a structured TLM algorithm, allowing for arbitrary placement of the panel between the model nodes. In the present paper the single- and multi-layered thin sheet models described in [15] are applied to multi-layered CFC panels; the technique is further extended in two important ways. The first of these is to include material anisotropy and the second is to incorporate the panel model within a three-dimensional unstructured TLM (UTLM) code, again using a piece-wise linear approach. The latter advance is commensurate with our recent work to incorporate material losses within the UTLM codes [17].

The paper is structured as follows. In the next section the 1D TLM model of a uniform medium is briefly introduced in order to provide a basis for the further discussions. In Section III important concepts regarding the implementation of the 1D CFC model are reviewed to aid the reader, initially for a single layer and then for multiple layers. The further extension to allow for material anisotropy is then introduced. Results presented in Section IV establish the accuracy and stability of the 1D TLM CFC models. Section V describes the new implementation of the 2D thin CFC surface model within a three-dimensional (3D) unstructured TLM code, its validation and subsequent application to a realistic 3D geometry. Finally some conclusions are drawn in Section VI.

## II. 1D TLM Model In the Uniform Medium

The TLM is a numerical differential time-domain method based on the analogy between the field quantities and lumped circuit equivalents [7]. The method discretizes the modeling space using a mesh of transmission lines, connected at nodes. The field is represented using the voltage wave pulses which propagate and scatter through the mesh at every time step. The schematic presentation of the lossless 1D TLM node is given in Fig.1 where a mesh node is defined by a total voltage $V_n$. Materials with properties different from free space are modeled by introducing stubs with the desired dielectric or magnetic behavior to the model.

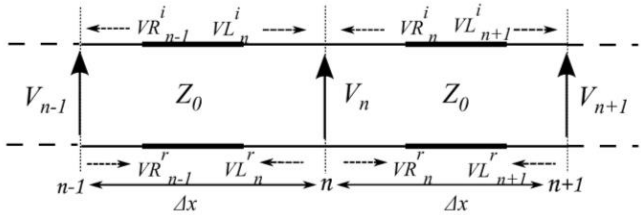

Fig1. Two sections of lossless transmission line connected at node $n$.

The TLM method, can be algorithmically separated into several stages namely; initialization, calculation of the voltages at all nodes, scattering and connection [7]. Initialization defines the sources and initial wave values; scattering determines the reflected voltage waves at all nodes and connection exchanges the voltage waves between adjacent nodes. Referring to Fig.1, this means that $VL_n^i$ and $VR_n^i$ are the incident voltages from the left and right of the node $n$, and $VL_n^r$ and $VR_n^r$ are the reflected voltages from the left and right of the node $n$, respectively.

The total voltage at node $n$ at time $k$ is,

$$_kV_n = {_kVL_n^i} + {_kVR_n^i}. \quad (1)$$

The reflected voltages can be obtained from the incident voltages through the following scattering process

$$\begin{aligned} _kVL_n^r &= {_kV_n} - {_kVL_n^i}, \\ _kVR_n^r &= {_kV_n} - {_kVR_n^i}. \end{aligned} \quad (2)$$

At the next time step, $k+1$, the reflected voltages will become the incident voltages of the adjacent nodes through the connection process,

$$\begin{aligned} _{k+1}VR_n^i &= {_kVL_{n+1}^r}, \\ _{k+1}VL_n^i &= {_kVR_{n-1}^r}. \end{aligned} \quad (3)$$

Following an initial excitation, these two processes are repeated at each node for the desired number of time steps to obtain the required solutions.

## III. 1D CFC Model in TLM

This section outlines the implementation of the 1D CFC model within the TLM mesh. Embedding of the thin CFC panel within the TLM mesh is schematically given in Fig.2, where $V_1^i$ and $V_1^r$ are the incident and reflected voltage of port 1, and $V_2^i$ and $V_2^r$ are the incident and reflected voltage of port 2, respectively.

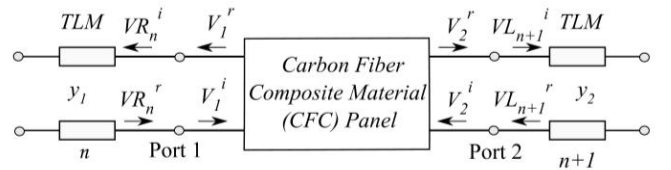

Fig.2 CFC panel embedded between TLM nodes

The embedding of the thin panel requires that at each time step the reflected voltages on both ports need to be solved both in terms of incident voltages and in terms of CFC panel characteristics.

### A. Single layered CFC model

A thin CFC panel with thickness $d$ can be viewed as a transmission line of length $d$ as shown in Fig.3.

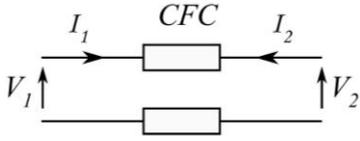

Fig.3 The transmission line model of CFC panel

Its admittance matrix is given by [18]

$$\begin{pmatrix} I_1 \\ I_2 \end{pmatrix} = -j \begin{pmatrix} Y\cot\theta & -Y\csc\theta \\ -Y\csc\theta & Y\cot\theta \end{pmatrix} \begin{pmatrix} V_1 \\ V_2 \end{pmatrix}, \quad (4)$$

where $\theta = \omega d\sqrt{LC}$ is the electrical length and $Y = \sqrt{C/L}$ the admittance of the CFC panel, and both $L$ and $C$ can be complex to account for losses.

When it is embedded between TLM nodes as in Fig.2, both ends of this CFC panel are driven by Thévenin equivalent circuits. As shown in [15] the total voltages can, after some manipulation, be expressed by

$$\begin{aligned}&\left(y_1 y_2 + YY - jY(y_1+y_2)\cot\theta\right)\begin{pmatrix}V_1\\V_2\end{pmatrix}\\ &= \begin{pmatrix} y_2 - jY\cot\theta & -jY\csc\theta \\ -jY\csc\theta & y_1 - jY\cot\theta \end{pmatrix}\begin{pmatrix} 2y_1 V_1^i \\ 2y_2 V_2^i \end{pmatrix}.\end{aligned} \quad (5)$$

The cotangent and cosecant functions in (5) can be expanded as infinite summations [19] as shown in equations (6) and (7). These infinite expansions can be numerically truncated after N terms; the value of $N$ will be explored in Section III.

$$jY\cot\theta = j\sqrt{\frac{C}{L}}(\frac{1}{\theta} + 2\theta\sum_{k=1}^{N=\infty}\frac{1}{\theta^2 - k^2\pi^2}), \quad (6)$$

$$jY\csc\theta = j\sqrt{\frac{C}{L}}(\frac{1}{\theta} + 2\theta\sum_{k=1}^{N=\infty}\frac{(-1)^k}{\theta^2 - k^2\pi^2}), \quad (7)$$

Bilinear Z-transformation [20] and digital filter theory [21] are then used to transfer the equations from the frequency domain to the time domain, as in [15]. This transforms (5) into the form (8),

$$\sum_i \frac{P_i(z)}{Q_i(z)} y(z) = \sum_i \frac{R_i(z)}{S_i(z)} x(z), \quad (8)$$

where $P_i(z), Q_i(z), R_i(z)$ and $S_i(z)$ are the first order or second order polynomials in $z$, $x$ and $y$ are the input and the output of the system.

Each term $R_i(z)/S_i(z)$ on the right hand side of (8) can be seen as the transfer function of a digital filter. Then the summation of all the terms can be viewed as the parallel combination of a number of first and second order digital filters.

The overall output of the right side of (6), $w(n\Delta t)$, can be obtained through summing the output of each filter together,

$$w(n\Delta t) = \sum_i w_i(n\Delta t). \quad (9)$$

Therefore, equation (8) becomes,

$$\sum_i \frac{P_i(z)}{Q_i(z)} y(z) = w(z). \quad (10)$$

Assuming

$$u_i(z) = \frac{P_i(z)}{Q_i(z)} y(z), \quad (11)$$

and considering $Q_i(z) = Q_{i0} + Q_{i1}z^{-1} + Q_{i2}z^{-2}$, equation (11) can be written as

$$u_i(z) = \frac{P_i(z)}{Q_{i0}} y(z) - (\frac{Q_i(z)}{Q_{i0}} - 1)u_i(z). \quad (12)$$

Summing both sides of (12) yields,

$$\left(\sum_i \frac{P_i(z)}{Q_{i0}}\right) y(z) = \sum_i (\frac{Q_i(z)}{Q_{i0}} - 1)u_i(z) + w(z). \quad (13)$$

After applying an inverse Z transform [20] to both sides of (13), the final output in the time domain, $y(n\Delta t)$, can be obtained.

*B. Multilayered CFC model*

In this section a three layered CFC panel is taken as an example to demonstrate the embedding of a multilayered CFC model.

The transmission line model of the embedded three layered CFC panel is shown in Fig.4. The admittance matrix of each layer can be written as,

$$\begin{pmatrix} I_1 \\ I_2 \end{pmatrix} = \begin{pmatrix} y_1 - jY_1\cot\theta_1 & jY_1\csc\theta_1 \\ jY_1\csc\theta_1 & -jY_1\cot\theta_1 \end{pmatrix} \cdot \begin{pmatrix} V_1 \\ V_2 \end{pmatrix}, \quad (14)$$

$$\begin{pmatrix} -I_2 \\ I_3 \end{pmatrix} = \begin{pmatrix} -jY_2\cot\theta_2 & jY_2\csc\theta_2 \\ jY_2\csc\theta_2 & -jY_2\cot\theta_2 \end{pmatrix} \cdot \begin{pmatrix} V_2 \\ V_3 \end{pmatrix}, \quad (15)$$

$$\begin{pmatrix} -I_3 \\ I_4 \end{pmatrix} = \begin{pmatrix} -jY_3\cot\theta_3 & jY_3\csc\theta_3 \\ jY_3\csc\theta_3 & y_2 - jY_3\cot\theta_3 \end{pmatrix} \cdot \begin{pmatrix} V_3 \\ V_4 \end{pmatrix}, \quad (16)$$

where $Y_1, Y_2$, and $Y_3$ are the admittances of the first, second and third layer of this CFC panel.

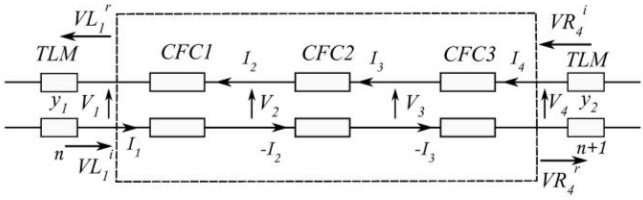

Fig.4 Transmission line model of multilayered CFC panel

Combining (14), (15), and (16), the following linear matrix equations can be obtained,

$$\begin{pmatrix} I_1 \\ 0 \\ 0 \\ I_2 \end{pmatrix} = \begin{pmatrix} V_1 \\ V_2 \\ V_3 \\ V_4 \end{pmatrix} \cdot$$

$$\begin{pmatrix} y_1 - jY_1\cot\theta_1 & jY_1\csc\theta_1 & 0 & 0 \\ jY_1\csc\theta_1 & -jY_1\cot\theta_1 - jY_2\cot\theta_2 & jY_2\csc\theta_2 & 0 \\ 0 & jY_2\csc\theta_2 & -jY_3\cot\theta_3 - jY_2\cot\theta_2 & jY_3\csc\theta_3 \\ 0 & 0 & jY_3\csc\theta_3 & y_2 - jY_3\cot\theta_3 \end{pmatrix}$$

(17)

Considering that

$$I_1 = 2y_1 \cdot VL_1^i, I_4 = 2y_2 \cdot VR_4^i,  \qquad (18)$$

the terms on the left side of (17) are known and the matrix can be solved for unknown voltages. In the present work this linear equation set has been solved using the iterative matrix solver based on the Gauss-Seidel method [20]. Typically for a CFC panel with *n* layers, there are *n* admittance matrices, resulting in a square matrix of order (*n+1*) with (*n+1*) unknowns [15].

### C. Extension to anisotropic materials

The electric properties of anisotropic materials vary in different directions [22]. Here, an anisotropic CFC panel is modeled. The panel is regarded as being anisotropic with respect to two in-plane coordinates (*x* and *y*); the principal directions of the anisotropy are assumed to be orthogonal to each other and to align with the x and y directions. The response of a flat CFC panel to an incident plane wave whose electric field is polarized along one of the coordinate directions can be modeled locally using a 1D CFC model as described in part A of this section. In the case of an arbitrary polarization, the electric field can be written as

$$\vec{E}^i = E_x \cdot \hat{x} + E_y \cdot \hat{y} = |E^i|\cos\varphi \cdot \hat{x} + |E^i|\sin\varphi \cdot \hat{y}, \qquad (19)$$

where φ is the angle between the electric field and the *x* axis, and $\hat{x}$ and $\hat{y}$ are the in plane unit vectors. The incident field is first decomposed into components aligning with the anisotropic principal axes. Each of these components excites a different 1D CFC model and the net reflection and transmission is be determined by combining the two responses [23].

## IV. NUMERICAL RESULTS, 1D CASE

In this section the 1D CFC models derived are used to study the reflection and transmission characteristics and shielding effectiveness of CFC panels. Both isotropic and anisotropic single layered CFC panels are considered, in subsections A and B respectively. In sub-section C the multilayered CFC model is also validated through two examples.

### A. Single Layered CFC panels: Isotropic case

An isotropic single layered CFC panel having a thickness of 1 mm, dielectric constant $\varepsilon_r = 4.56$ and conductivity $\sigma_e = 8000 S/m$ is considered, as in [24]. The TLM mesh size is set to 0.01 m and a delta pulse is used to initialize the problem. The reflection and transmission waves in the time domain can be obtained through the model proposed, and in the frequency domain using the Fast Fourier transform (FFT).

The infinite summations used to represent the cotangent and cosecant functions can be expanded as an infinite summation and must be truncated for computational purposes. In this example, the order of the expansions, *N*, used in equations (6) and (7), was chosen to be 10, 20, and 100 to test the convergence of the model.

Fig. 5 shows the reflection ($S_{11}$) and transmission ($S_{21}$) coefficients of the single layer CFC panel for different orders of expansion *N*, and compares them with the analytical ones obtained from a network approach [25] in the frequency domain.

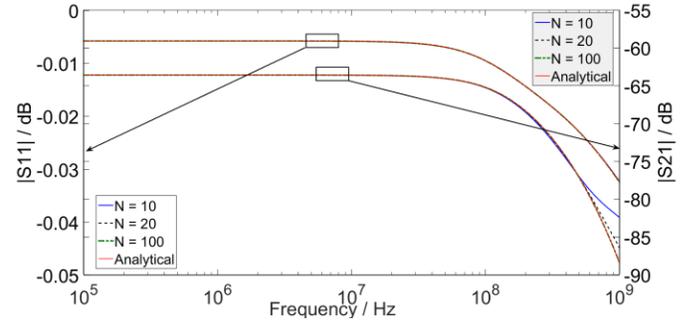

Fig.5 Reflection and Transmission coefficients of single layered CFC panel with 1 mm thickness and $\varepsilon_r = 4.56, \sigma_e = 8000S/m$. *N* is the order of the expansions used for the cotangent and cosecant functions that appear in equations (6) and (7).

Fig.5 shows that the agreement between numerical and analytical results for $S_{11}$ is excellent regardless of the number of expansion terms used in the range presented. The results for $S_{21}$ decrease with the order of the expansion especially in the high frequency region. When *N* = 100, the numerical results of both reflection and transmission coefficients agree with the analytical ones over a wide frequency span. The number of terms used to get perfect results depends on the parameters of the panel. The runtime for *N* = 100 takes 0.375s using a PC with an Intel Core 2 Duo CPU 3GHz processor and 4GB memory.

### B Single Layered CFC panels: Anisotropic case.

An anisotropic CFC panel whose electrical properties in the $x$ direction are $\varepsilon_{rx} = 4.56, \sigma_{ex} = 8 \times 10^3 S/m$ [24] and in the $y$ direction are $\varepsilon_{rx} = 2, \sigma_{ex} = 10^4 S/m$ [1] is considered. The thickness of the panel is 1 mm. Fig.6 shows the reflection and transmission coefficients of this anisotropic CFC panel as a function of the incident angle at a frequency of 1 GHz. The order of the expansions, $N$, used is 50, which was found to be sufficient to produce results having very good agreement with the analytical ones, as demonstrated in the Figure.

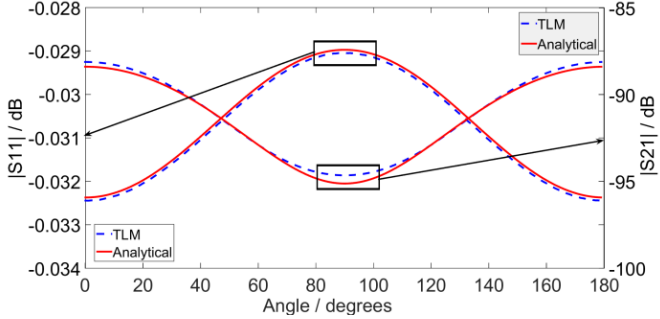

Fig.6. Reflection ($S_{11}$) and Transmission ($S_{22}$) coefficients of an anisotropic CFC panel against the incident angle at 1 GHz

*C. Multilayered CFC panels*

The first example used to test the accuracy of the multilayered CFC model is a symmetric three layered CFC panel. The first and third layers of this panel have the same electric properties, $\varepsilon_r = 4.56, \sigma_e = 8 \times 10^3 S/m$ [24], and the second layer has a material with $\varepsilon_r = 2, \sigma_e = 10^4 S/m$ [1]. All the layers have the same thickness of 1 mm.

Fig.7 shows the reflection and transmission coefficients of this three layered CFC panel as a function of frequency. The numerical results with different expansion terms $N$ are compared with the analytical ones obtained from a network approach [25] in the frequency domain.

It can be seen that all the numerical results show good agreement with the analytical ones, with the agreement becoming closer when more terms are used to approximate the infinite series for the cotangent and cosecant functions. The runtime for $N$ = 50 is 0.4s.

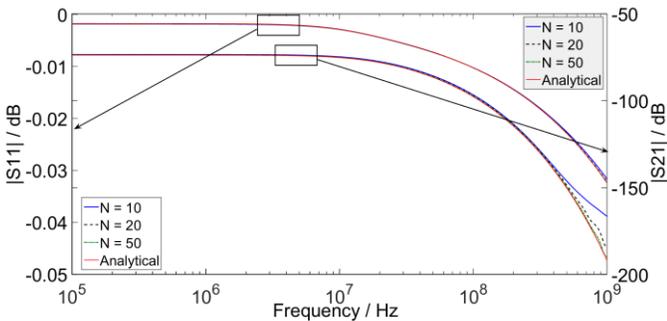

Fig.7 Reflection ($S_{11}$) and Transmission ($S_{22}$) coefficients of a symmetric three layered CFC panel, in which the first and third layers have the materials of $\varepsilon_r = 4.56, \sigma_e = 8 \times 10^3 S/m$, and the second layer with materials of $\varepsilon_r = 2, \sigma_e = 10^4 S/m$. All layers have a thickness of 1 mm.

As a further application, the multilayered CFC model developed was used to calculate the shielding effectiveness of four different layered CFC panels, which were originally studied in [1]. The shielding effectiveness to the electric field, $SE_E$ is defined as in [1]

$$SE_E = 20\log(E_i / E_t), \quad (20)$$

where $E_i$ and $E_t$ are the incident and transmitted electric field respectively.

The properties of the four kinds of CFC panels studied are given in Table I [1].

TABLE I
PARAMETERS OF MULTILAYERED CFC PANELS

| Panel | No. of layers | Layer conductivity(S/m) | Layer relative permittivity | Layer thickness(mm) |
|---|---|---|---|---|
| A | 1 | $\sigma_1 = 10^4$ | $\varepsilon_{r1} = 2$ | $d_1 = 1$ |
| B | 3 | $\sigma_1 = 10^4$<br>$\sigma_2 = 50$<br>$\sigma_3 = 10^3$ | $\varepsilon_{r1} = 2$<br>$\varepsilon_{r2} = 4$<br>$\varepsilon_{r3} = 3$ | $d_1 = 0.6$<br>$d_2 = 0.6$<br>$d_3 = 0.6$ |
| C | 5 | $\sigma_1 = \sigma_3 = 10^4$<br>$\sigma_2 = \sigma_4 = 50$<br>$\sigma_5 = 10^3$ | $\varepsilon_{r1} = \varepsilon_{r3} = 2$<br>$\varepsilon_{r2} = \varepsilon_{r4} = 4$<br>$\varepsilon_{r5} = 3$ | $d_1 = d_3 = 0.2$<br>$d_2 = d_4 = 0.2$<br>$d_5 = 0.2$ |
| D | 9 | $\sigma_1 = \sigma_3 = \sigma_8 = 10^4$<br>$\sigma_2 = \sigma_4 = \sigma_6 = 50$<br>$\sigma_5 = \sigma_7 = \sigma_9 = 10^3$ | $\varepsilon_{r1} = \varepsilon_{r3} = \varepsilon_{r8}$<br>$= 2$<br>$\varepsilon_{r2} = \varepsilon_{r4} = \varepsilon_{r6}$<br>$= 4$<br>$\varepsilon_{r5} = \varepsilon_{r7} = \varepsilon_{r9}$<br>$= 3$ | $d_1 = d_3 = d_8 = 0.2$<br>$d_2 = d_4 = d_6 = 0.2$<br>$d_5 = d_7 = d_9 = 0.2$ |

Fig.8 shows the shielding performances of these four panels as calculated using the present method (TLM) and compares them with the FDTD results from [1]. It can be observed that the two sets of results show excellent agreement.

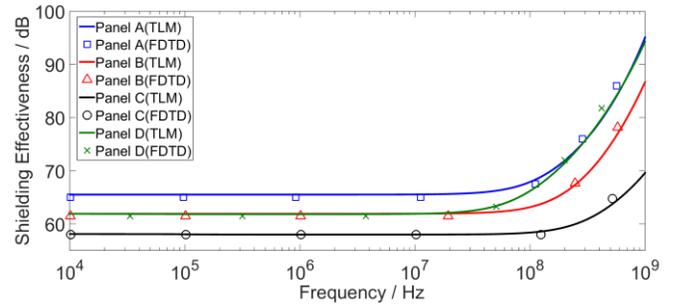

Fig.8 Shielding performances of the four panels whose properties are listed in Table I

## V. APPLICATION TO EMBEDDED 2D SURFACES

Having established the accuracy and stability of the 1D TLM CFC model, attention is now focused upon its deployment in the practical case of 2D CFC surfaces embedded within a 3D domain, for example the skin of an aircraft wing.

In 3D, the TLM method has been long established for Cartesian grids comprising cuboidal cells and more recently an unstructured mesh variant, UTLM, has been successfully demonstrated that operates on a tetrahedral mesh [11] – [14]. The use of an unstructured mesh permits piecewise linear approximations to curved surfaces and a straightforward grading of mesh size in the proximity of fine features. Therefore, UTLM permits the simulation of complex geometries such as aircraft structures with relative ease. However, thin panels of CFC materials are a common occurrence and typically their small thickness with respect to the overall problem size still demands special treatment.

Physically, the penetration of electromagnetic fields through thin CFC layers is a quasi-1D behavior in many cases. The reasonable assumption underlying this proposition is that there is little transverse diffraction of the fields compared with the dominant diffusive behavior in the direction through the panel. This is supported by the high levels of conductivity of CFCs, see for example Table I, such that, as shown in Fig. 8, significantly over 50 dB of shielding effectiveness is not untypical.

Accepting the 1D local assumption, and considering the canonical equivalent network of the interconnection between adjacent UTLM cells shown in Fig. 9, it is straightforward to insert the digital filter model developed in section 3 for the 2D simulations, including multi-layer and anisotropic effects, between the pairs of adjacent tetrahedral cells whose interface is defined to contain the thin panel. It is noted that in this model, the thin panel does not occupy any physical space in the mesh. This is advantageous for meshing, although there are situations where this may be inconvenient for maintaining the relative positioning of different parts of the structure. Moreover, for certain particular geometrical configurations, such as CFC joints and apertures, the locally 1D behavior assumption of the model may not be sufficiently justified. However, there is no reason why a designer cannot choose to return to an explicitly meshed representation for these cases with the appropriate material parameters being used in standard UTLM cells.

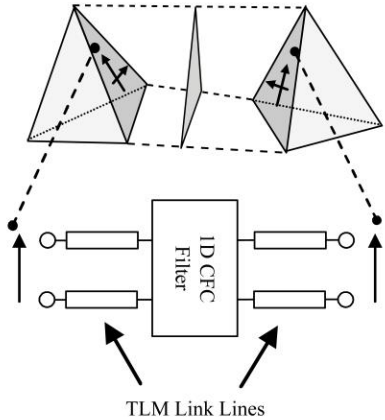

Fig 9. The canonical scattering cell in the UTLM method. The CFC filters are inserted between the link lines connecting the faces on adjacent cells, one for each polarization (only one shown here).

*A. Validation*
In order to validate the model for 3D problems within UTLM, the resonant frequencies of a sphere of radius 1m made from 1 mm thick single-layer CFC panel were calculated using the UTLM method for various mesh configurations. It is re-emphasized that the CFC panel is not meshed but embedded between the adjacent nodes by using the model described above. The properties of the CFC materials were chosen as $\varepsilon_r = 2, \sigma_e = 10^4 \, S/m$ [1]. The geometry was meshed for the UTLM simulations using in-house Delaunay Mesher software. Figure 10 shows the meshed structure with the maximum surface mesh area of 0.002 m$^2$. As shown in the figure, several complementary spheres were built in order to calculate the resonant frequencies of the sphere. The problem is excited by a magnetic dipole source on the surface of the center sphere as shown in the figure. The magnetic field inside the CFC sphere was calculated by summing and averaging the magnetic fields on the observation sphere. A 4th order Butterworth filter with 0.5dB points of 100 MHz and 300 MHz is used to reduce the noise in the results. The resonant frequencies of the CFC sphere were obtained by Fast Fourier Transforming the calculated magnetic fields.

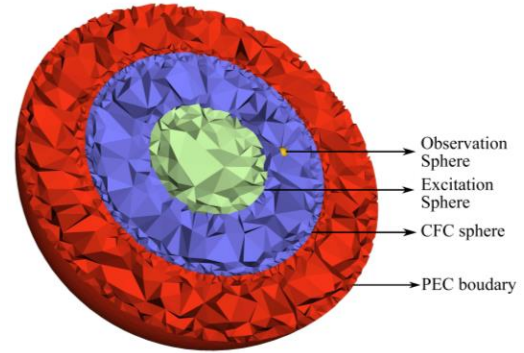

Fig. 10. The geometry of a CFC sphere meshed for UTLM simulations with maximum surface mesh area of 0.002 m$^2$.

In Figure 11, the UTLM results obtained are compared with the analytic solution for a metal sphere of the same radius obtained from [25] where the relative differences in the resonant frequencies between the CFC sphere and the equivalent metal sphere are shown against the maximum surface mesh area. The first six TE and TM mode resonant frequencies are convergent as the mesh size becomes smaller. The relative differences are within 1%, which validates the embedded CFC model.

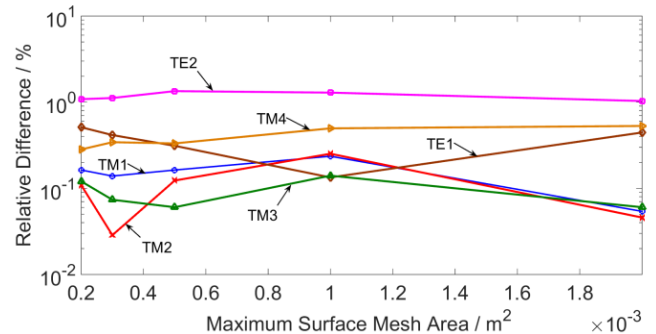

Fig. 11. The relative differences in the resonant frequencies between the CFC sphere and the equivalent metal sphere.

*B. Application to a 3D CFC Wing*

To demonstrate the application of the 2D surface model within a 3D problem, Figure 12 shows the result of scattering a plane wave from a wing made from a 1 mm thick CFC skin. The CFC parameters are $\varepsilon_r = 4.56, \sigma_e = 8 \times 10^3 \, S/m$ [24], realized with an 8th order model in the simulation ($N = 8$ in equations (6) and (7)). The wing geometry was generated from an NACA 0025 airfoil [26] cross-section of chord length 2 m which was also rotated around its center to create the rounded wingtip. This geometry was then sampled using the Marching Tetrahedral Technique [27] and a 3D hybrid cuboidal-tetrahedral Delaunay mesh generated with in-house Delaunay Mesher software.

An incident plane wave was launched from a Huygens' surface on the outer box of the domain. This traveled in the x direction with the electric field vector polarized along the wing, i.e. the z direction. The time step for simulation was 8.3 ps, consistent with that typically used for the background cubic mesh and a total time of 66.4 ns was simulated. The plane wave was modulated by a Gaussian time envelope with a 3dB cut-off frequency of 2 GHz. The magnitude of the tangential magnetic fields was observed on the surface of the wing and on a pair of slices taken through the mesh. These fields have been passed through a 4th order Butterworth filter with 3dB points of 90 and 110 MHz and Fig. 12 shows a qualitative time snap shot after 29.1 ns of the simulation.

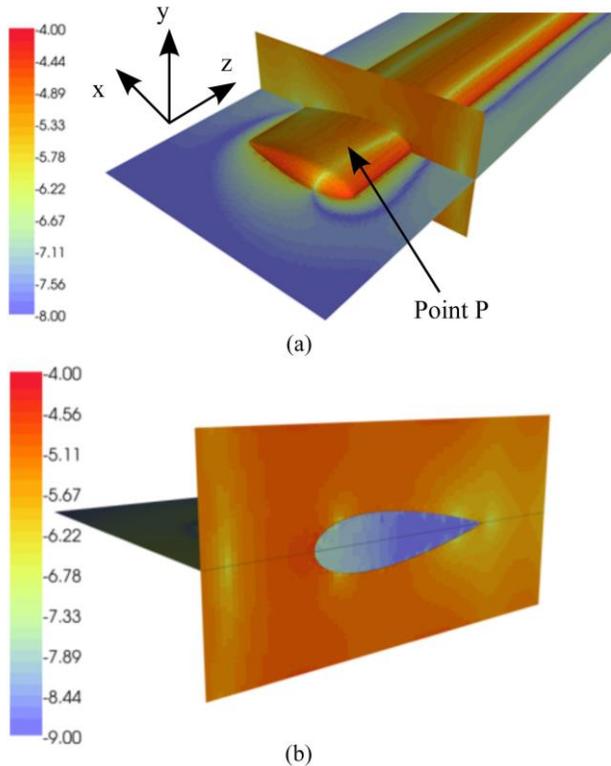

Fig. 12 The magnitude of the tangential magnetic fields in dB on the outside of the wing and a pair of field slices. This field profile occurs after 29 ns of the simulation and has been bandpass filtered centered on 100 MHz.

In the figure, it is clear that the field penetration into the inside of the wing is present and small and that the wing scatters causes cross-polarization scattering of the incident magnetic field from $H_y$ to $H_x$ and $H_z$. Fig. 13 (a) shows the magnitude of the tangential electric field recorded just inside the wing relative to the value recorded just outside the wing at the point P marked in Fig.12. In this example, the frequency domain results shown in Figure 13 (b) are virtually the same as those predicted by simple transmission through an infinite extent CFC sheet as shown in Fig.5, considering the approximate relation, $E_{inside}/E_{outside} = S_{21}/(1+S_{11})$, where the resonant effects inside the cavity are not considered.

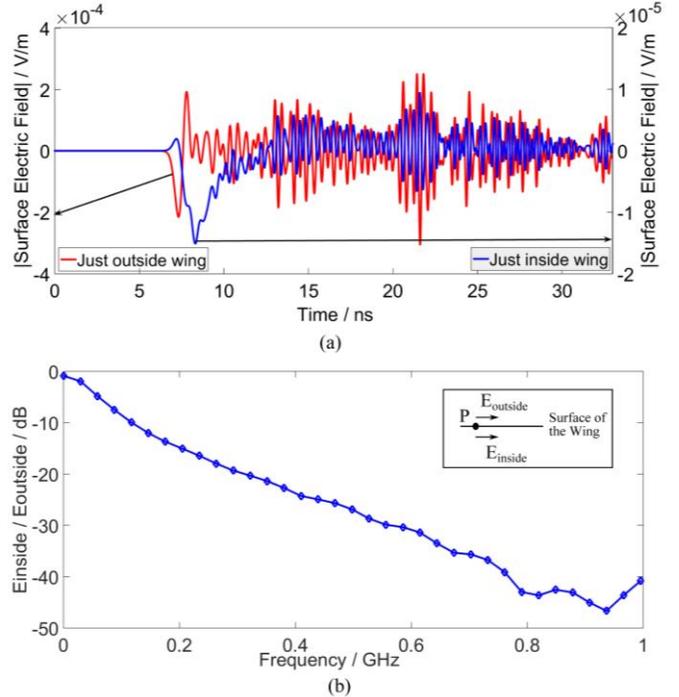

Fig. 13 Measures of the field penetration into the wing. (a) The time evolution of the tangential electric fields measured on either side of the CFC skin at point P in Fig. 12 and (b) the relative magnitude in of these measures against frequency.

VI. CONCLUSION

A new TLM model for single and multiple layered thin panels is developed based on Z transform and digital filter theory. It has also been extended to model anisotropic thin panels. In this model, thin panels are not meshed but embedded between two adjacent nodes, allowing for a relatively large mesh size in the simulation thus saving computational costs. Single- and multi-layered CFC panels and anisotropic CFC panels are taken as examples to test the convergence and accuracy of the model. When more terms are used in the expansions, the numerical results are shown to converge to the analytical ones.

With the validation of the 1D thin panel model, the 2D thin surface model embedded into 3D UTLM codes was introduced for the first time. It was validated by comparing the resonant frequencies of the CFC sphere and the equivalent metal sphere, and applied to model the scattering of a CFC wing with the plane wave illuminated. These examples show the applicability and capability of the thin panel model in the complex practical problems.

ACKNOWLEDGMENT

X. Meng would like to thank the China Scholarship Council for their financial support.

The work outlined above was carried out as part of ICE-NITE (see http://www.liv.ac.uk/icenite/), a collaborative research project supported by Innovate UK under contract reference 101665. The project consortium includes BAE Systems Limited (coordinator), Bombardier, MIRA Limited, Transcendata Europe Limited, The University of Liverpool, and The University of Nottingham.

The authors wish to thank Prof. C. Jones and Dr. S. Earl of BAE systems for valuable discussions.